\documentstyle[prl,aps,epsfig]{revtex}
\begin{document}

\title{Phase Structure of Nambu--Jona-Lasinio Model at Finite Isospin Density }
\author{ Lianyi He and Pengfei Zhuang\\
        Physics Department, Tsinghua University, Beijing 100084, China }
\date{\today}
\maketitle

\begin{abstract}
In the frame of flavor $SU(2)$ Nambu--Jona-Lasinio model with
$U_A(1)$ breaking term we found that, the structure of two chiral
phase transition lines does not exist at low isospin density in
real world, and the critical isospin chemical potential for pion
superfluidity is exactly the pion mass in the vacuum.
\end{abstract}

\noindent {\bf pacs numbers:} 11.10.Wx,\ \ 12.38.-t\ \ 25.75.Nq

\section {Introduction}
It is generally believed that there exists a rich phase structure
of Quantum Chromodynamics (QCD) at finite temperature and baryon
density, for instance, the deconfinement process from hadron gas
to quark-gluon plasma, the transition from chiral symmetry
breaking phase to the symmetry restoration phase\cite{dandc}, and
the color superconductivity\cite{csc} at low temperature and high
baryon density. Recently, the study on the QCD phase structure is
extended to finite isospin density. The physical motivation to
study isospin spontaneous breaking and the corresponding pion
superfluidity is related to the investigation of compact stars,
isospin asymmetric nuclear matter and heavy ion collisions at
intermediate energies.

While the perturbation theory of QCD can describe well the
properties of the new phases at high temperatures and/or high
densities, the study on the phase structure at moderate (baryon or
isospin) density depends on lattice QCD calculation and effective
models with QCD symmetries. While there is not yet precise lattice
result at finite baryon density due to the Fermion sign
problem\cite{sign}, it is in principle no problem to do lattice
simulation at finite isospin density\cite{CB1}. It is
found\cite{LA} that the critical isospin chemical potential for
pion condensation is about the pion mass in the vacuum, $\mu_I^c
\simeq m_\pi$. The QCD phase structure at finite isospin density
is also investigated in many low energy effective models, such as
chiral perturbation theory\cite{CB1,CB2,CB3}, Nambu--Jona-Lasinio
(NJL) model\cite{NJL1,NJL2,NJL3},ladder QCD\cite{ladder}, strong
coupling lattice QCD\cite{str} and random matrix
method\cite{random}.

One of the models that enables us to see directly how the dynamic
mechanisms of chiral symmetry breaking and restoration operate is
the NJL model\cite{NJL4} applied to quarks\cite{NJL5}. Within this
model, one can obtain the hadronic mass spectrum and the static
properties of mesons remarkably well\cite{NJL5,NJL6}, and the
chiral phase transition line\cite{NJL5,NJL6,NJL7} in the
temperature and baryon chemical potential ($T-\mu_B$) plane is
very close to the one calculated with lattice QCD. Recently, this
model is also used to investigate the color superconductivity at
moderate baryon density\cite{NJL8,NJL9,NJL10,NJL11,NJL12}. In the
study at finite isospin density, it is predicted\cite{NJL1} in
this model that the chiral phase transition line in $T-\mu_B$
plane splits into two branches. This phenomena is also found in
random matrix method\cite{random} and ladder QCD\cite{ladder}.
Since the NJL Lagrangian used in \cite{NJL1} does not contain the
determinant term which breaks the $U_A(1)$ symmetry and leads to
reasonable meson mass splitting, it is pointed out \cite{NJL3}
that the presence of the $U_A(1)$ breaking term will cancel the
structure of the two chiral phase transition lines, if the
coupling constant describing the $U_A(1)$ breaking term is large
enough. However, this coupling constant is considered as a free
parameter and not yet determined in \cite{NJL3}. Another problem
in the NJL calculation at finite isosipn density is that the
critical isospion chemical potential for pion condensation
$\mu_I^c = m_\pi$ is not recovered in the model\cite{NJL2}. In
this letter we will focus on these two problems in the frame of
NJL model with $U_A(1)$ breaking term. We hope to derive the
critical isospin chemical potential exactly and try to fix the
chiral structure by fitting the meson masses in the vacuum.

The letter is organized as follows. We present at finite
temperature and baryon and isospin densities the chiral and pion
condensates in mean field approximation and meson masses in Random
Phase Approximation (RPA) in Section 2, determine the coupling
constant of the $U_A(1)$ breaking term in Section 3, and then
analytically prove the relation $\mu_I^c = m_\pi$ in Section 4. We
conclude in Section 5.

\section {NJL model at finite temperature and baryon and isospin densities}
We start with the flavor $SU(2)$ NJL model defined by
\begin{equation}
\label{njl1}
{\cal L}
=\bar\psi(i\gamma^\mu\partial_\mu-m_0+\mu\gamma_0)\psi+{\cal
L}_{int}\ ,
\end{equation}
where $m_0$ is the current quark mass, $\mu$ the chemical
potential matrix in flavor space,
\begin{equation}
\label{mu} \mu=\left(\begin{array}{cc}\mu_u & 0 \\ 0 &
\mu_d\end{array}\right) = \left(\begin{array}{cc}{\mu_B\over
N_c}+{\mu_I\over 2} & 0\\ 0 & {\mu_B\over N_c}-{\mu_I\over
2}\end{array}\right)
\end{equation}
with $\mu_B$ and $\mu_I$ being the baryon and isospin chemical
potential, respectively, and the interaction part
includes\cite{NJL3} the normal four Fermion couplings
corresponding to scalar mesons $\sigma, a_0, a_+$ and $a_-$ and
pseudoscalar mesons $\eta\prime, \pi_0, \pi_+$ and $\pi_-$
excitations, and the 't-Hooft\cite{Hooft} determinant term for
$U_A(1)$ breaking,
\begin{eqnarray}
\label{njl2}
{\cal L}_{int}
&=&\frac{G}{2}\sum_{a=0}^{3}\left[(\bar{\psi}\tau_a\psi)^{2}
+(\bar{\psi}i\gamma_{5}\tau_a\psi)^{2}\right]+
\frac{K}{2}\left[\det\bar{\psi}(1+\gamma_{5})\psi+\det\bar{\psi}(1-\gamma_{5})\psi\right]\nonumber\\
&=&\frac{1}{2}(G+K)\left[(\bar{\psi}\psi)^{2}+(\bar{\psi}i\gamma_{5}\vec{\tau}\psi)^{2}\right]
+\frac{1}{2}(G-K)\left[(\bar{\psi}\vec{\tau}\psi)^{2}+(\bar{\psi}i\gamma_{5}\psi)^{2}\right]\
.
\end{eqnarray}
For $K=0$ and $m_0=0$, the Lagrangian is invariant under $
U_B(1)\bigotimes U_A(1) \bigotimes SU_V(2) \bigotimes SU_A(2)$
transformations, but for $K\neq 0$, the symmetry is reduced to
$U_B(1) \bigotimes SU_V(2) \bigotimes SU_A(2)$ and the $U_A(1)$
breaking leads to $\sigma$ and $a$ mass splitting and $\pi$ and
$\eta\prime$ mass splitting. If $G=K$, we come back to the
standard NJL model\cite{NJL5} with only $\sigma, \pi_0, \pi_+$ and
$\pi_-$ mesons.

We introduce the quark condensates
\begin{equation}
\label{sigma1}
\sigma_u = \langle\bar u u\rangle \ ,\ \ \ \
\sigma_d = \langle\bar d d\rangle\ ,
\end{equation}
or equivalently the $\sigma$ and $a_0$ condensates
\begin{eqnarray}
\label{sigma2}
\sigma &=& \langle\bar\psi\psi\rangle = \langle\bar
u u+\bar d d\rangle = \sigma_u +\sigma_d
\ ,\nonumber\\
a_0 &=&\langle\bar\psi\tau_3\psi\rangle = \langle\bar u u-\bar d
d\rangle = \sigma_u - \sigma_d\ ,
\end{eqnarray}
and the pion condensate
\begin{eqnarray}
\label{pion}
{\pi\over \sqrt 2} &=& \langle\bar{\psi} i\gamma_5
\tau_+\psi\rangle = \langle\bar{\psi} i\gamma_5\tau_-\psi\rangle =
{1\over \sqrt 2}\langle\bar{\psi}i\tau_1\gamma_5\psi\rangle\ ,
\end{eqnarray}
where we have chosen the pion condensate to be real. The quark
condensate and pion condensate are, respectively, the order
parameter of chiral phase transition and pion superfluidity. By
separating each Lorentz scalar in the Lagrangian (\ref{njl2}) into
the classical condensate and the quantum fluctuation, and keeping
only the linear terms in the fluctuations, one obtains the
Lagrangian in mean field approximation,
\begin{equation}
\label{njl3} {\cal L}_{mf} = \bar\psi{\cal
S}_{mf}^{-1}\psi-G(\sigma_u^2+\sigma_d^2)-2K\sigma_u\sigma_d-{G+K\over
2}\pi^2\ ,
\end{equation}
where ${\cal S}_{mf}^{-1}$ is the inverse of the mean field quark
propagator, in momentum space it reads
\begin{equation}
\label{s-1} {\cal S}_{mf}^{-1}(k)=\left(\begin{array}{cc}
\gamma^\mu k_\mu+\mu_u\gamma_0-M_u &
i(G+K)\pi\gamma_5\\
i(G+K)\pi\gamma_5 & \gamma^\mu
k_\mu+\mu_d\gamma_0-M_d\end{array}\right)
\end{equation}
with the effective quark masses
\begin{eqnarray}
\label{qmass}
M_u &=& m_0-2G\sigma_u-2K\sigma_d\ ,\nonumber\\
M_d &=& m_0-2G\sigma_d-2K\sigma_u\ .
\end{eqnarray}

The thermodynamic potential of the system in mean field
approximation can be expressed in terms of the effective quark
propagator,
\begin{equation}
\label{omega} \Omega(T,\mu_B,\mu_I;\sigma_u,\sigma_d,\pi) =
G(\sigma_u^2+\sigma_d^2)+2K\sigma_u\sigma_d+{G+K\over
2}\pi^2-\frac{T}{V}\ln \det{{\cal S}^{-1}_{mf}(k)}\ .
\end{equation}

The condensates $\sigma_u, \sigma_d$ and $\pi$ as functions of
temperature and baryon and isospin chemical potentials are
determined by the minimum thermodynamic potential,
\begin{equation}
\label{minimum}
\frac{\partial \Omega}{\partial \sigma_u}=0,\ \ \
\frac{\partial \Omega}{\partial \sigma_d}=0, \ \ \ \frac{\partial
\Omega}{\partial \pi}=0\ .
\end{equation}
It is easy to see from the chemical potential matrix and the quark
propagator matrix that for $\mu_B=0$ or $\mu_I=0$ the gap
equations for $\sigma_u$ and $\sigma_d$ are symmetric, and one has
\begin{eqnarray}
\label{sigma3}
&& \sigma_u = \sigma_d = {\sigma\over 2}\ ,\ \ \ \ \ \ \ a_0 = 0\ ,\nonumber\\
&& M_u = M_d = M_q = m_0 -(G+K)\sigma\ .
\end{eqnarray}

The quark propagator in flavor space can be formally expressed as
\begin{equation}
\label{s1}
{\cal S}_{mf}(k)= \left(\begin{array}{cc} {\cal S}_{uu}(k)&{\cal S}_{ud}(k)\\
{\cal S}_{du}(k)&{\cal S}_{dd}(k)\end{array}\right)\ .
\end{equation}
From the comparison with the definition of the condensates
(\ref{sigma1}) and (\ref{pion}), one can express the condensates
in terms of the matrix elements of ${\cal S}_{mf}$,
\begin{eqnarray}
\label{gap1}
\sigma_u &=& -\int\frac{d^4k}{(2\pi)^4} {\bf Tr}\left[i{\cal S}_{uu}(k)\right]\ ,\nonumber\\
\sigma_d &=& -\int\frac{d^4k}{(2\pi)^4} {\bf Tr}\left[i{\cal
S}_{dd}(k)\right]\ ,\nonumber\\
\pi &=& \int\frac{d^4k}{(2\pi)^4} {\bf Tr}\left[\left({\cal
S}_{ud}(k)+{\cal S}_{du}(k)\right)\gamma_5\right]\ .
\end{eqnarray}
where the trace is in color and spin space.

In the self-consistent mean field approximation, it is well known
that the meson masses $M_M$ as the bound states of the colliding
quark-antiquark pairs are determined as the poles of the meson
propagators in RPA at zero momentum,
\begin{equation}
\label{pole1}
1-(G+K)\Pi_M\left(k_0=M_M,\bf k=0\right)=0
\end{equation}
for $\sigma$ and $\pi$, and
\begin{equation}
\label{pole2}
1-(G-K)\Pi_M\left(k_0=M_M,\bf k=0\right)=0
\end{equation}
for $\eta\prime$ and $a$, where $\Pi_M$ is the meson polarization
function
\begin{equation}
\label{polari1} -i\Pi_M(k) = -\int{d^4p\over (2\pi)^4} {\bf
Tr}\left[\Gamma_M^* i{\cal S}_{mf}(p+k)\Gamma_M i{\cal
S}_{mf}(p)\right]\ ,
\end{equation}
and $\Gamma_M$ and $\Gamma_M^*$ are the interaction vertexes
\begin{equation}
\label{vertex}
\Gamma_M = \left\{\begin{array}{ll}
1 & M=\sigma\\
\tau_3 & M=a_0 \\
\tau_+ & M=a_+ \\
\tau_- & M=a_-\\
i\gamma_5 & M=\eta^\prime\\
i\gamma_5\tau_3 & M=\pi_0 \\
i\gamma_5\tau_+ & M=\pi_+ \\
i\gamma_5\tau_- & M=\pi_-
\end{array}\right.\ ,\ \ \ \ \ \
\Gamma_M^* = \left\{\begin{array}{ll}
1 & M=\sigma\\
\tau_3 & M=a_0 \\
\tau_- & M=a_+ \\
\tau_+ & M=a_-\\
i\gamma_5 & M=\eta^\prime\\
i\gamma_5\tau_3 & M=\pi_0 \\
i\gamma_5\tau_- & M=\pi_+ \\
i\gamma_5\tau_+& M=\pi_-
\end{array}\right.
\end{equation}
with $\tau_\pm = (\tau_1 \pm i\tau_2)/\sqrt 2$. Doing the trace in
flavor and color spaces, one has
\begin{eqnarray}
\label{polari2} && \Pi_{a_+}(k) = 2N_ci\int {d^4p\over (2\pi)^4}
{\bf tr}\left[
{\cal S}_{uu}(p+k){\cal S}_{dd}(p)\right]\ ,\\
&& \Pi_{a_-}(k) = 2N_ci\int {d^4p\over (2\pi)^4} {\bf tr}\left[
{\cal
S}_{dd}(p+k){\cal S}_{uu}(p)\right]\ ,\nonumber\\
&& \Pi_{\pi_+}(k) = -2N_ci\int {d^4p\over (2\pi)^4} {\bf tr}\left[
\gamma_5{\cal S}_{uu}(p+k)\gamma_5{\cal S}_{dd}(p)\right]\ ,\nonumber\\
&& \Pi_{\pi_-}(k) = -2N_ci\int {d^4p\over (2\pi)^4} {\bf tr}\left[
\gamma_5{\cal
S}_{dd}(p+k)\gamma_5{\cal S}_{uu}(p)\right]\ ,\nonumber\\
&& \Pi_\sigma (k) = \Pi_{a_0}(k) = N_ci\int {d^4p\over (2\pi)^4}
{\bf tr} \left[{\cal S}_{uu}(p+k){\cal S}_{uu}(p)+{\cal
S}_{ud}(p+k){\cal S}_{du}(p)+{\cal S}_{du}(p+k){\cal
S}_{ud}(p)+{\cal
S}_{dd}(p+k){\cal S}_{dd}(p)\right]\ ,\nonumber\\
&& \Pi_{\eta^\prime} (k) = \Pi_{\pi_0}(k)\nonumber\\
&& = -N_ci\int {d^4p\over (2\pi)^4} {\bf tr}\left[\gamma_5{\cal
S}_{uu}(p+k)\gamma_5{\cal S}_{uu}(p)-\gamma_5{\cal
S}_{ud}(p+k)\gamma_5{\cal S}_{du}(p)-\gamma_5{\cal
S}_{du}(p+k)\gamma_5{\cal S}_{ud}(p)+\gamma_5{\cal
S}_{dd}(p+k)\gamma_5{\cal S}_{dd}(p)\right]\ ,\nonumber
\end{eqnarray}
now the trace is taken only in spin space.

\section {Chiral phase structure }
We now consider the QCD phase structure below the minimum isospin
chemical potential $\mu_I^c$ for pion superfluidity. Since the
pion condensate is zero, there is only chiral phase structure in
this region.

The simple diagonal matrix ${\cal S}_{mf}^{-1}(k)$ in this region
makes it easy to calculate the matrix elements of the effective
quark propagator, they can be expressed explicitly as
\begin{eqnarray}
\label{element}
{\cal S}_{uu}(k) &=&
\frac{\Lambda_+^u\gamma_0}{k_0-E_1}+
\frac{\Lambda_-^u\gamma_0}{k_0+E_2}\ ,\nonumber\\
{\cal S}_{dd}(k) &=& \frac{\Lambda_+^d\gamma_0}{k_0-E_3}+\frac{\Lambda_-^d\gamma_0}{k_0+E_4} ,\nonumber\\
{\cal S}_{ud}(k) &=& {\cal S}_{du}(k) =0\ ,
\end{eqnarray}
with quasiparticle energies
\begin{eqnarray}
\label{energy1}
&& E_1 = E_u-\mu_u\ ,\ \ \ \ E_2 = E_u+\mu_u\ ,\ \
\ \
E_3 = E_d-\mu_d\ ,\ \ \ \ E_4 = E_d+\mu_d\ ,\nonumber\\
&& E_u =\sqrt{{\bf k}^2+M_u^2}\ ,\ \ \ \ E_d=\sqrt{{\bf
k}^2+M_d^2}\ ,
\end{eqnarray}
and energy projectors
\begin{equation}
\label{project}
\Lambda_{\pm}^{u,d} = {1\over
2}\left(1\pm{\gamma_0\left({\bf \gamma\cdot k}+M_{u,d}\right)\over
E_{u,d}}\right)\ .
\end{equation}
After performing the Matsubara frequency summation in
(\ref{gap1}), the gap equations determining the two quark
condensates as functions of temperature and baryon and isospin
chemical potentials read
\begin{eqnarray}
\label{gap2} \sigma_u &=& -2N_c\int\frac{d^3{\bf
k}}{(2\pi)^3}\frac{M_u}{\sqrt{k^2+M_u^2}}(1-f(E_1)-f(E_2))\ ,
\nonumber\\
\sigma_d &=& -2N_c\int\frac{d^3{\bf
k}}{(2\pi)^3}\frac{M_d}{\sqrt{k^2+M_d^2}}(1-f(E_3)-f(E_4))\ ,
\end{eqnarray}
with the Fermi-Dirac distribution function
\begin{equation}
f(x) = {1\over e^{x/T} + 1}\ .
\end{equation}

We now first discuss the real case, $N_c=3$. It is well known that
in the absence of isospin degree of freedom the temperature and
baryon density effects on chiral symmetry restoration are
different\cite{NJL5}: The chiral condensate drops down
continuously with increasing temperature, which means a second
order phase transition, but jumps down suddenly at a critical
baryon density, which indicates a first order phase transition. At
finite baryon density and finite isospin density, while the
density behavior of the $u$- and $d$-quark condensates are
different to each other, they may jump down at the same critical
baryon chemical potential or at two different critical points. In
the case without considering the $U_A(1)$ breaking term, the two
critical points do exist, and therefore, there are two chiral
phase transition lines\cite{NJL1} in the temperature and baryon
chemical potential plane at fixed isospin chemical potential. What
is the effect of the $U_A(1)$ breaking term on the QCD phase
structure? The calculation in the NJL model showed\cite{NJL3} that
if there exists a structure of two chiral phase transition lines
depends on the coupling constant $K$ of the determinant term. Let
\begin{equation}
\label{alpha1} \alpha = {1\over 2}\left(1-{G-K\over
G+K}\right)={K\over G+K}\ ,
\end{equation}
it was found\cite{NJL3} that at $\mu_I = 60$ MeV the two-line
structure disappears for $\alpha > 0.11$. This means that the
two-line structure is true only at small ratio $\alpha$. However,
$\alpha$ is a free parameter in \cite{NJL3} and the two coupling
constants $G$ and $K$ are not yet separately determined.

Before the discussion of the problem in real world with nonzero
current quark mass, it is instructive to analyze the phase
structure in chiral limit with $m_0 =0$ even though pion
condensation will happen at finite $\mu_I$. One can regard this as
the limit of $m_0\rightarrow0$. From the gap equations
(\ref{gap2}) one can clearly see that when one of the quark
condensates becomes zero, the other one is forced to be zero for
any coupling constants $G$ and $K\ne 0$. Therefore, there is only
one chiral phase transition line at any $K\ne 0$ in chiral limit.
The two-line structure happens only at $K =0$. In this case, the
two gap equations decouple, the two chiral phase transition lines
are determined by
\begin{eqnarray}
\label{chiralgap} && 1-12G\int{d^3{\bf k}\over (2\pi)^3}{1\over
|{\bf k}|}\left[1-f\left(|{\bf k}|-{\mu_B\over 3}-{\mu_I\over
2}\right)-f\left(|{\bf k}|+{\mu_B\over 3}+{\mu_I\over
2}\right)\right]=0\
,\nonumber\\
&& 1-12G\int{d^3{\bf k}\over (2\pi)^3}{1\over |{\bf
k}|}\left[1-f\left(|{\bf k}|-\left({\mu_B\over
3}-\mu_I\right)-{\mu_I\over 2}\right)-f\left(|{\bf
k}|+\left({\mu_B\over 3}-\mu_I\right)+{\mu_I\over
2}\right)\right]=0\ ,
\end{eqnarray}
the difference in baryon chemical potential between the two
critical points at fixed $T$ and $\mu_I$ is
\begin{equation}
\label{chiralgap2} \Delta\mu_B(T,\mu_I) = 3\mu_I\ .
\end{equation}

In real world there are four parameters in the NJL model, the
current quark mass $m_0$, the three-momentum cutoff $\Lambda$, and
the two coupling constants $G$ and $K$. Among them $m_0, \Lambda$
and the combination $G+K$ can be determined by fitting the chiral
condensate $\sigma$, the pion mass $m_\pi$ and the pion decay
constant $f_\pi$ in the vacuum. In \cite{NJL3} $m_0 = 6$ MeV,
$\Lambda = 590$ MeV, and $(G+K)\Lambda^2/2 = 2.435$, for which
$\sigma = 2(-241.5$ MeV$)^3$, $m_\pi = 140.2$ MeV, and $f_\pi =
92.6$ MeV. To determine the two coupling constants separately or
the ratio $\alpha$, one needs to know the $\eta\prime$- or
$a$-meson properties in the vacuum.

In the vacuum the two quark condensates degenerate, and the pole
equations determining the meson masses are much simplified as
\begin{eqnarray}
\label{pole3} &&1 - 4N_c(G+K)\int{d^3{\bf k}\over (2\pi)^3}{1\over
E_q}\frac{E_q^2-M_q^2}{E_q^2-m_\sigma^2/4}= 0\ ,\nonumber\\
&&1 - 4N_c(G-K)\int{d^3{\bf k}\over (2\pi)^3}{1\over
E_q}\frac{E_q^2-M_q^2}{E_q^2-m_a^2/4}= 0\ ,\nonumber\\
&&1 - 4N_c(G+K)\int{d^3{\bf k}\over (2\pi)^3}{1\over
E_q}\frac{E_q^2}{E_q^2-m_\pi^2/4}= 0\ ,\nonumber\\
&&1 - 4N_c(G-K)\int{d^3{\bf k}\over (2\pi)^3}{1\over
E_q}\frac{E_q^2}{E_q^2-m_{\eta^\prime}^2/4}= 0\ ,
\end{eqnarray}
with quark energy $E_q = \sqrt{{\bf k}^2+M_q^2}$. It is easy to
see that if $K=0$, the masses of $\sigma-$ and $a$-mesons
degenerate and the masses of $\eta^\prime$- and $\pi$-mesons
degenerate, and if $K=G$ as in the standard NJL Lagrangian, the
$a$- and $\eta\prime$-mesons disappear in the model.

The mass equation for $\pi$-meson can be used to determine the
combination $G+K$, which is already considered in \cite{NJL3}, and
the mass equation for $\eta\prime$-meson is related to the
combination $G-K$ and then to the ratio
\begin{equation}
\label{alpha2} \alpha = {1\over 2}\left(1-{1\over G+K}{1\over 12
\int{d^3{\bf k}\over (2\pi)^3}{E_q\over
E_q^2-m_{\eta\prime}^2/4}}\right)\ .
\end{equation}
Combining with the above known parameters $m_0, \Lambda$ and $G+K$
obtained in \cite{NJL3} and choosing $m_{\eta\prime} = 958$ MeV,
we have $\alpha =0.29$ which is much larger than the critical
value $0.11$\cite{NJL3} for the two-line structure at $\mu_I = 60$
MeV. In fact, for a wide mass region $540$ MeV $<m_{\eta\prime}<
1190$ MeV, we have $\alpha > 0.11$, and there is no two-line
structure.

To fully answer the question if the two-line structure exists
before the pion condensation happens, we should consider the limit
$\mu_I = \mu_I^c$ where the difference in baryon chemical
potential between the two critical points is the maximum, if the
two-line structure happens. From the approximate result of lattice
simulation\cite{LA} and the exact result of NJL analyze in next
section, we know that the critical isospin chemical potential for
pion superfluidity is equal to the pion mass in the vacuum,
$\mu_I^c = m_\pi$. From the analysis above, at this value the
isospin chemical potential difference between the two critical
points corresponding, respectively, to $\sigma_u = 0$ and
$\sigma_d =0$ in chiral limit and with $K=0$ is $\Delta\mu_B =
3m_\pi$. In real world with $m_0 \neq 0$ and $K\neq 0$, with
increasing coupling constant $K$ or the ratio $\alpha$ the two
lines approach to each other and finally coincide at about $\alpha
= 0.21$ which is still less than the value $0.29$ calculated by
fitting $m_{\eta\prime} = 958$ MeV. Therefore, the two-line
structure disappears if we choose $m_{\eta\prime} = 958$ MeV. In
fact, for $720$ MeV $< m_{\eta\prime} <  1140$ MeV, we have still
$\alpha > 0.21$ and the two phase transition lines are cancelled
in this wide mass region. When the $\eta\prime$-meson mass is
outside this region, the $U_A(1)$ breaking term is not strong
enough to cancel the two-line structure, but the two lines are
already very close to each other. Considering we have used the
$\eta^\prime$ mass in three flavor world, we should use a smaller
$\eta^\prime$ mass in two flavor world. Using the approximate
relation $m_{\eta^\prime}^2\propto N_f$ one can estimate
$m_{\eta^\prime}\approx780MeV$ which is in the region where
$\alpha
> 0.21$.

At this stage one may ask in which case the two chiral phase
transition lines at finite isospin chemical potential appear. To
answer this question we pay attention to the relation between the
$\eta^\prime$ mass and $N_c$. It is believed the $\eta^\prime$
mass decreases with increasing $N_c$\cite{UA1,UA2}. At
sufficiently large $N_c$, the $U_A(1)$ breaking induced flavor
mixing effect can be neglected and there will be two chiral phase
transition lines.

\section {Critical isospin chemical potential for pion condensation}
Since both thermal motion and nonzero baryon number will increase
the critical isospin density for pion superfluidity, the minimum
isospin chemical potential $\mu_I^c$ corresponds to the parameters
$T=0$ and $\mu_B = 0$. In this case the pion condensate is zero
and the two quark condensates degenerate, the pole equations for
the meson masses are then reduced to
\begin{eqnarray}
\label{pole3} && 1-4N_c(G+K)\int{d^3{\bf k}\over (2\pi)^3}{1\over
E_q}{E_q^2-M_q^2\over E_q^2-M_\sigma^2/4}=0\
,\nonumber\\
&& 1-4N_c(G-K)\int{d^3{\bf k}\over (2\pi)^3}{1\over
E_q}{E_q^2-M_q^2\over E_q^2-M_{a_0}^2/4}=0\
,\nonumber\\
&& 1-4N_c(G-K)\int{d^3{\bf k}\over (2\pi)^3}{1\over
E_q}{E_q^2-M_q^2\over E_q^2-\left(M_{a_\pm}\pm\mu_I\right)^2/4}=0\
,\nonumber\\
&& 1-4N_c(G+K)\int{d^3{\bf k}\over (2\pi)^3}{1\over
E_q}{E_q^2\over E_q^2-M_{\pi_0}^2/4}=0\
,\nonumber\\
&& 1-4N_c(G+K)\int{d^3{\bf k}\over (2\pi)^3}{1\over
E_q}{E_q^2\over E_q^2-\left(M_{\pi_\pm}\pm\mu_I\right)^2/4}=0\
,\nonumber\\
&& 1-4N_c(G-K)\int{d^3{\bf k}\over (2\pi)^3}{1\over
E_q}{E_q^2\over E_q^2-M_{\eta\prime}^2/4}=0\ ,
\end{eqnarray}
where the quark mass $M_q$ satisfies the same gap equation as in
the vacuum,
\begin{equation}
\label{gap5} M_q-m_0-4N_c(G+K)M_q \int{d^3{\bf k}\over (2\pi)^3}
{1\over E_q} = 0\ ,
\end{equation}
namely,
\begin{equation}
\label{mq0} M_q(T=0,\mu_B=0,\mu_I\le\mu_I^c) = M_q(0,0,0)\ .
\end{equation}
Therefore, from the comparison with the meson mass equations
(\ref{pole2}) in the vacuum, we derive the isospin dependence of
the meson masses for $\mu_I \le \mu_I^c$,
\begin{eqnarray}
\label{mass2}
&& M_\sigma(\mu_I) = m_\sigma\ ,\nonumber\\
&& M_{a_0}(\mu_I) = m_a\ ,\nonumber\\
&& M_{a_\pm}(\mu_I) = m_{a_\pm}\mp\mu_I\ ,\nonumber\\
&& M_{\pi_0}(\mu_I) = m_\pi\ ,\nonumber\\
&& M_{\pi_\pm}(\mu_I) = m_{\pi_\pm}\mp\mu_I\ ,\nonumber\\
&& M_{\eta\prime}(\mu_I) = m_{\eta\prime}\ .
\end{eqnarray}
We see that the mesons with zero isospin charge keep their vacuum
values, the mesons with positive isospin charge drop down linearly
in the isospin chemical potential, and the mesons with negative
isospin charge go up linearly in the isospin chemical potential.
Since $m_a > m_\pi$, the above isospin dependence will firstly
break down at $\mu_I = m_\pi$. Beyond this value the mass of
$\pi_+$ becomes negative and it makes no sense. This gives us a
strong hint that the end point of the normal phase without pion
condensation or the starting point of the pion superfluidity phase
is at $\mu_I^c = m_\pi$.

To prove this hint we should consider directly the isospin
behavior of the order parameter of the pion superfluidity, namely
the pion condensate. Deriving the matrix elements ${\cal S}_{ud}$
and ${\cal S}_{du}$ from (\ref{s-1}) and then performing the
Matsubara frequency summation, the gap equation determining
self-consistently the pion condensate as a function of isospin
chemical potential at $T=0$ and $\mu_B = 0$ reads,
\begin{equation}
\label{gap4} \pi\left[1-2N_c(G+K)\int{d^3{\bf k}\over
(2\pi)^3}\left({1\over
\sqrt{\left(E_q-\mu_I/2\right)^2+\left(G+K\right)^2\pi^2}}+{1\over
\sqrt{\left(E_q+\mu_I/2\right)^2+\left(G+K\right)^2\pi^2}}\right)\right]=0\
.
\end{equation}
Obviously, the minimum isospin chemical potential $\mu_I^c$ where
the pion superfluidity starts is controlled by the square bracket
with $\pi =0$,
\begin{equation}
\label{critical} 1-4N_c(G+K)\int{d^3{\bf k}\over (2\pi)^3}{1\over
E_q}{E_q^2\over E_q^2-\left(\mu_I^c\right)^2/ 4} = 0\ .
\end{equation}
From the comparison of (\ref{critical}) with the pion mass
equation (\ref{pole2}) in the vacuum, we find explicitly that the
critical isospin chemical potential $\mu_I^c$ for the pion
superfluidity phase transition at $T=\mu_B=0$ is exactly the
vacuum pion mass $m_\pi$,
\begin{equation}
\mu_I^c = m_\pi\ .
\end{equation}
We emphasize that this conclusion is independent of the model
parameters and the regularization scheme, it is a general
conclusion in mean field approximation for quarks and RPA
approximation for mesons. Unlike the chiral phase structure which
strongly depends on $U_A(1)$ breaking coupling $K$ and color
degree of freedom $N_c$, the critical isospin chemical potential
$\mu_I^C$ does not depend on them.

\section {Conclusions}
We have investigated the two flavor NJL model with $U_A(1)$
breaking term at finite isospin density, as well as at finite
temperature and baryon density. With a self-consistent treatment
of quarks in mean field approximation and mesons in RPA, we
determined the two coupling constants separately by fitting the
meson masses in the vacuum, and then investigated the phase
structure for chiral symmetry restoration and pion superfluidity,
especially the isospin effect on the chiral phase transition line
and the minimum isospin chemical potential for pion superfluidity.

The main conclusions are:\\
{\bf 1)} The two chiral phase transition lines in the $T-\mu_B$
plane predicated by the NJL model without $U_A(1)$ breaking term
is cancelled by the strong $U_A(1)$ breaking term at low isospin
chemical potential in real world ($N_c=3$). Only when the isospin
chemical potential $\mu_I$ is close to the critical value
$\mu_I^c$ of pion condensation, there is probably the two-line
chiral structure, depending on the $\eta\prime$-meson mass.
Therefore, in relativistic heavy ion collisions where the typical
$\mu_I$ value is much less than $\mu_I^c$, it looks impossible to
realize the
two-line structure. However, it can be realized in large $N_c$ QCD.\\
{\bf 2)} The critical isospin chemical potential for pion
condensation in NJL model is exactly the pion mass in the vacuum,
$\mu_I^c = m_\pi$, independent of the model parameters, the
regularization scheme, the $U_A(1)$ breaking term and the color
degree of freedom $N_c$.

The temperature and baryon and isospin density dependence of the
chiral and pion condensates as well as the meson masses, and the
extension to flavor $SU_3$ NJL model are under way.

{\bf Acknowledgments:} We thank Dr. Meng Jin for helpful
discussions. The work was supported in part by the grants
NSFC10135030, SRFDP20040003103 and G2000077407.

\end{document}